\documentclass[%
reprint,
nofootinbib,
amsmath,amssymb,
aps,
onecolumn
]{revtex4-2}

\usepackage{graphicx}
\usepackage{dcolumn}
\usepackage{bm}
\usepackage{color}
\usepackage{comment}
\usepackage{hyperref}
\usepackage{url}
\usepackage{amsmath}

\begin{document}

\title{Hamiltonian formulation of the supersymmetric KdV equation}

\author{Ali Pazarci}
 \affiliation{Department of Physics, Bogazici University,
34342 Bebek, Istanbul, Turkey}

\author{Nadir Ghazanfari}
\affiliation{Department of Electrical and Electronics Engineering, Istinye University,
34485 Istanbul, Turkey}

\author{Ilmar Gahramanov}
\affiliation{Department of Physics, Bogazici University,
34342 Bebek, Istanbul, Turkey}%
\affiliation{Center for Mathematics and its Apllications, Khazar University,
Mehseti St. 41, AZ1096, Baku, Azerbaijan}

\date{\today}

\begin{abstract}
     We studied the constrained Hamiltonian formulation of a supersymmetric Korteweg–de Vries (KdV) equation, which is observed to be a constrained system similar to its classical version. We found a nontrivial Lagrangian description, where we select $a=2$ for the free parameter $a$ in the supersymmetric extension. The corresponding degenerate Lagrangian requires an exclusive consideration and the utilization of the Dirac–Bergmann algorithm. We explicitly determined the full set of primary and secondary constraints and constructed the total Hamiltonian governing the dynamics of the system. In this analysis, in addition to a nontrivial constraint involving the fermionic fields, the consistency conditions give rise to a nonlocal contribution to the Hamiltonian density. This highlights a distinctive feature of this supersymmetric extension. We showed that the resulting Hamilton equations of motion reproduce the supersymmetric KdV system in the component form. Finally, we derived a compact superspace representation of the Hamiltonian and demonstrated its consistency with the component-level formulation.
\end{abstract}
\maketitle
\hfill {\textbf{\textit{Dedicated to the memory of Yavuz Nutku }}}


\section{Introduction}

The Korteweg-de Vries (KdV) equation is a completely integrable nonlinear partial differential equation~\cite{fadeev1971a, Flaschka1991, Newellch4,sonja2024} and provides a rich area of study involving analytical methods, geometric structures, and physical applications~\cite{Kruskal1965,dubrovin19740a1,Whithamch16,Huang2005,Donagi_2003,Nicolas2018}. This equation is known for its soliton solutions~\cite{Lax1968,marchenko0a,dubrovin19740a2,Dickey2003,Ma2022,ma2024}, infinite number of independent conserved quantities~\cite{Miki1976,nutku_1985}, and multi-Hamiltonian structure~\cite{ANTONOWICZ198795,nutku_1987,nutku1990,nutku-gumral-1990,Nutku2001,Nutku2002}. The study of the KdV equation has led to developments in different fields of physics, ranging from wave analysis (shallow water, acoustic, and electromagnetic waves)~\cite{Morikawa1969,Zabusky_Galvin_1971,lannes2013,Verheest_Hereman_2025} to inverse scattering theory~\cite{Gardner-Kruskal-Miura-1967,Zakharov-Shabat-1974,nutku-gurses-1981,grudsky2023,rybkin2026}.

An interesting extension of integrable systems like KdV arises from the supersymmetric generalizations of these equations by introducing fermionic degrees of freedom. The inclusion of fermionic fields often reveals algebraic and geometric structures that do not emerge in purely bosonic systems. In this context, different formulations of the supersymmetric KdV (sKdV-a) system have been developed, which are often a family of equations characterized by a real parameter $a$, with distinct properties depending on its value. For example, only for some values of $a$, the integrability and supersymmetry coexist in a nontrivial way for the sKdV-a equation.

The classical theory of the inverse variational problem, namely, the reconstruction of a Lagrangian from a (or a set of) prescribed equation(s) of motion, is well understood for ordinary differential equations through the Helmholtz conditions and related constructions such as the Jacobi last multiplier \cite{Nucci_2008,prince2021,montesinos2026}. However, its extension to nonlinear partial differential equations, especially in the presence of fermionic fields, remains largely unexplored. Moreover, even when a Lagrangian formulation exists, sometimes the consistent Hamiltonian is not derived from a local Lagrangian. Instead, it is constructed from the system's nature as an integrable hierarchy. On the other hand, the Lagrangian might be degenerate, which leads to a constrained Hamiltonian system.


In the case of the sKdV-a equation, a Lagrangian formulation exists only for specific values of the parameter $a$. Among these, the case $a = 0$ is trivial, as the system reduces to the ordinary KdV equation. A nontrivial Lagrangian formulation is available for the case $a = 2$, however, it is linear in the time derivatives of all fields and is therefore degenerate. Consequently, the Hamiltonian cannot be obtained through a straightforward Legendre transformation and instead requires a systematic constrained analysis using the Dirac–Bergmann algorithm (DBA).

In this article, we develop a complete constrained Hamiltonian formulation of the sKdV-2 system. Starting from the Lagrangian, we apply the DBA, explicitly identify the full set of primary and secondary constraints, and construct the total Hamiltonian. A nontrivial constraint relating the fermionic fields emerges from the consistency conditions. Furthermore, the consistency requirements of the constraints generate nonlocal contributions to the Hamiltonian density. We show that the resulting Hamiltonian equations reduce to the component form of the sKdV-2 system. We obtain a compact superspace representation of the Hamiltonian and demonstrate its consistency with previous works on this system.

Following the introduction given above, the paper is organized as follows: In Sec. II, we review the sKdV-a system and introduce the Lagrangian structure. In Sec. III, in detail, we construct the Hamiltonian expression in superspace using the generalized Dirac-Bergmann algorithm. Then, in Sec. IV, we discuss the case of the supersymmetric nonlinear Schrödinger equation and the difficulties that emerge in constructing a consistent Lagrangian. Finally, we conclude with a discussion of the results and possible directions for further investigation.

\section{Supersymmetric KdV system}
The first super-extension and the first supersymmetric formulation were introduced in \cite{Kupershmidt:1984pr}\footnote{Note that the super-extension of the KdV equation introduced in \cite{Kupershmidt:1984pr} is not supersymmetric, rather, it provides one of the earliest examples of integrability within a Grassmann algebraic framework.} and \cite{Manin:1985hu}, respectively. The supersymmetric KdV equation introduced in these works corresponds to the $N = 1$ case. Higher supersymmetric extensions of the KdV equation have also been investigated in the literature \cite{delduc1996,bonora1997,popowicz2019}. In this paper, we restrict our attention to the $N = 1$ sKdV equation.

 The supersymmetric generalization of the KdV equation 
\begin{equation}\label{kdvu}
    u_t + 6 u u_{x}+u_{3x}=0
\end{equation}
can be formulated by introducing a fermionic superfield~\cite{Mathieu1988}
\begin{equation}
    \Phi(x,t,\theta)= \xi(x,t) + \theta u(x,t),
\end{equation}
where $\xi(x,t)$ and $u(x,t)$ are fermionic and bosonic component fields, respectively, and $\theta$ is a real
Grassmann variable. A class of supersymmetric KdV equations is parametrized by a real parameter $a$, which we refer to as sKdV-a. This class, in terms of component fields, is given by
\begin{align}
    & \xi_t +(6-a)u \xi_x+a\xi u_x+\xi_{3x}=0,
    \\
    & u_t +6uu_x-a\xi\xi_{xx}+u_{3x}=0.
\end{align}
The structural properties of this supersymmetric system depend on the value of $a$~\cite{Mathieu1988}.  For example, the case $a=0$ corresponds to a trivial extension, for which one of the equations simply becomes the KdV equation itself. In addition, a Lagrangian formalism exists only for specific choices of $a$. However, for the case $a=2$, the system not only becomes nontrivial, but also has a consistent Lagrangian, which can be formulated.     

To establish a Hamiltonian formalism, it is convenient to introduce a velocity potential by defining $u\equiv u_x$. Accordingly, the KdV equation can be written as
\begin{equation}\label{kdvux}
    u_{xt} + 6 u_{x} u_{xx}+u_{4x}=0.
\end{equation}
The desired supersymmetric extension of \eqref{kdvux} in terms of the superfield $\Phi$ can be expressed as
\begin{equation}\label{skdvux}
    (D^2 \Phi)_t + a D^2(D^2 \Phi D^3\Phi)+(6-2a)D^3\Phi D^4\Phi +D^8\Phi=0,
\end{equation}
where $    D= \partial_\theta+\theta \partial_x $
is the supercovariant derivative for which $D^2=\partial_x$. Expanding \eqref{skdvux} in terms of the component fields gives the system
\begin{align}\label{skdv_xt1}
    & \xi_{xt} +(6-a)u_x \xi_{xx}+a\xi_x u_{xx}+\xi_{4x}=0,
    \\\label{skdv_xt2}
    & u_{xt} +6u_xu_{xx}-a\xi_x\xi_{3x}+u_{4x}=0.
\end{align}
\section{Hamiltonian Analysis}
We consider the variational formulation of sKdV-2, for which a Lagrangian density can be written as a function of both fermionic and bosonic component fields.
\begin{equation}\label{Lagrangian_skdv-2}
    \mathcal{L}_{sKdV-2}= -\frac{1}{2} u_xu_t+\frac{1}{2}\psi\psi_t-u_x^3-2u\psi\psi_{xx}+\frac{1}{2}u_{xx}^2-\xi_{xx}\psi_{xx}+\frac{1}{2}\xi_{xx}\xi_{3x}
\end{equation}
Here, $\psi(x,t)=\xi_x$ is a fermionic field introduced to write a consistent Lagrangian density. The Euler–Lagrange variations with respect to the fields $\phi=(u, \psi, \xi)$ generate a coupled system of equations, including \eqref{skdv_xt1} and \eqref{skdv_xt2}, in addition to
\begin{align}\label{constraint}
    \psi-\xi_x=0,
\end{align}
which arises from the modification of the Lagrangian density. Since the Lagrangian density \eqref{Lagrangian_skdv-2} is linear in the time derivatives of all fields in $\phi$, the determinant of the Hessian matrix vanishes.
\begin{equation}
    \left\vert
    \frac{\delta^2 \mathcal{L}}{\delta \dot{\phi}_i \delta \dot{\phi}_j} 
    \right\vert=0\,
\end{equation}
This means that the Lagrangian is degenerate (singular), and the system must be treated by using the DBA to systematically determine the constraints and construct a Hamiltonian formulation \cite{YNutku_1983,Nutku1984,filiz2018,gumral2022,Pazarci2023}. The total Hamiltonian $H$ is constructed by adding the contribution of the constraints, $\mathcal{H}_c$, to the canonical Hamiltonian density, $\mathcal{H}_L$.
\begin{equation}
    H = \int \mathcal{H}\:dx =\int (\mathcal{H}_L+\mathcal{H}_c)\:dx\,. 
\end{equation}
The canonical Hamiltonian is obtained from the Legendre transformation with the left derivative convention and can be expressed as
\begin{equation}
     \mathcal{H}_L= u_x^3+2u\psi\psi_{xx}-\frac{1}{2}u_{xx}^2+\xi_{xx}\psi_{xx}-\frac{1}{2}\xi_{xx}\xi_{3x}.
\end{equation}
According to DBA, the primary constraints are naturally chosen via the canonical momenta.
\begin{align}\label{momentas}
    & \Pi_u = - \frac{1}{2}u_x, \quad \Pi_\psi = - \frac{1}{2}\psi, \quad \Pi_\xi = 0,
\end{align}
conjugate to fields $u$, $\psi$, and $\xi$, respectively. Thus, the primary constraints can be defined as
\begin{align}
   c_1=\Pi_u + \frac{1}{2}u_x, \quad c_2=\Pi_\psi + \frac{1}{2}\psi, \quad c_3=\Pi_\xi\, .
\end{align}
The Lagrangian multipliers, $\lambda_i$'s, are introduced to construct the contribution of the constraints
\begin{equation}
    \mathcal{H}_c= \lambda_1 c_1+\lambda_2 c_2+\lambda_3 c_3,
\end{equation}
where $\lambda_2$ and $\lambda_3$ are Grassmann-odd variables.   

Consistent equations of motion can be constructed from these constraints only if they are constants of motion, a property that can be checked through the Poisson brackets of the constraints with the total Hamiltonian by using the following Poisson bracket relations between the fields and their conjugate momenta. 
\begin{align}
    \{u(x),\Pi_u(y)\}&=-\{\Pi_u(x),u(y)\}=\delta(x-y)
    \\
    \{\psi(x),\Pi_\psi(y)\}&=\{\Pi_\psi(x),\psi(y)\}=-\delta(x-y)
    \\
    \{\xi(x),\Pi_\xi(y)\}&=\{\Pi_\xi(x),\xi(y)\}=-\delta(x-y)
\end{align}
A secondary constraint arises from the consistency condition of the constraint $c_3$,
\begin{align}
    \{c_3,H\}=(\psi-\xi_x)_{4x}=0,
\end{align}
which leads to a secondary constraint,
\begin{equation}
    \tilde{c}_1 =\psi-\xi_x\,.
\end{equation}
The contribution of this constraint must also be added to the Hamiltonian density
\begin{equation}
    \mathcal{H}_c= \lambda_1 c_1+\lambda_2 c_2+\lambda_3 c_3 + \Tilde{\lambda}_1\Tilde{c}_1\, , 
\end{equation}
with a new Grassmann-odd Lagrange multiplier $\Tilde{\lambda}_1$. No further constraints are generated, therefore, the Lagrange multipliers can be determined through the consistency conditions, as listed below.
\begin{align}
    \lambda_1 &=2\psi \psi_x-3u_x^2-u_{3x} ,
    \\
    \lambda_2 &=-4u_x\psi_x-2\psi u_{xx}-\psi_{3x} ,
    \\
    \lambda_3 &= -2\partial_x^{-1}(u_x\psi_x)-2\psi u_x -\psi_{xx} ,
    \\
    \Tilde{\lambda}_1 &=\psi_{3x}-\xi_{4x}\, .
\end{align}

The total Hamiltonian density is completely constructed by substituting these expressions into the final constrained Hamiltonian. The total Hamiltonian density 
\begin{align}\label{total_hamiltonian_density}\nonumber
    \mathcal{H} &= \frac{1}{2}\xi_{xx}\xi_{3x}+\psi\psi_xu_x-\frac{1}{2} u_{x}^3+\frac{1}{2} \psi_x \psi_{xx}+\psi_{xx}\xi_{xx}+\Pi_u \left( 2\psi \psi_x -3 u_x^2-u_{3x}\right) 
    \\&
    +\Pi_\psi \left(4u_x\psi_x+2\psi u_{xx} +\psi_{3x} \right) +\Pi_\xi\left( 2 \partial_x^{-1}\left( u_x\psi_x\right)+2\psi u_x+\psi_{xx}\right),
\end{align}
contains a nonlocal term in the form of the inverse derivative operator appearing in $\lambda_3$~\cite{das-popowicz-2000}. This is a consequence of the constrained dynamics and arises from the fact that the fundamental physical variable here is $\xi_x$, not $\xi$. 

The time evolution of the variables is governed by the Hamilton equations
\begin{align}
    \dot{\phi}_i(x,t)= \{\phi_i,\phi_j\}\frac{\delta H}{\delta \phi_j},
\end{align}
where this time $\phi=(u,\Pi_u,\psi,\Pi_\psi,\xi,\Pi_\xi)$. The Hamilton equations of motion then, can be obtained as follows: 
\begin{align}
     & u_t - 2\psi\psi_x + 3u_x^2 + u_{3x}=0,
     \\&
     \psi_t + 4u_x\psi_x + 2\psi u_{xx} + \psi_{3x}=0,
     \\&
     \xi_t + 2 \partial_x^{-1}\left( u_x\psi_x\right) + 2\psi u_x + \psi_{xx}=0,
     \\&
     \psi_t + 4u_x\psi_x + 2\psi u_{xx} + 3\psi_{3x} - 2\xi_{4x}=0,
     \\&
     u_{xt} - 2\psi\psi_{xx} + 6 u_x u_{xx} + u_{4x}=0,
     \\&
     \left(\psi-\xi_x\right)_{4x}=0.
\end{align}
The Hamilton equations of motion lead to (\ref{skdv_xt1}) and (\ref{skdv_xt2}), up to some arbitrary functions of time that can be set to zero without loss of generality. In terms of the fermionic superfield $\Phi(x,t,\theta)$, the total Hamiltonian \eqref{total_hamiltonian_density} can be expressed as
\begin{align}\label{hamiltonian_in_superspace}
    \Bar{H}=\frac{1}{2}\int \left[-2 D^2 \Phi \left(D^3 \Phi\right)^2+D^4\Phi D^5 \Phi\right]\:dx d\theta\,.
\end{align}
Here, the overbar indicates that the Hamiltonian is written in superspace formulation. The relation between $\xi$ and $\psi$ lifts the nonlocality in the superspace expression of the Hamiltonian. The expression above is in agreement with the Hamiltonian obtained in earlier studies~\cite{Mathieu1988}, up to the identification $\Phi \equiv D^2\Phi$ due to the introduction of the velocity potential definition $u \equiv u_x$ in our analysis. Furthermore, in the limit where the fermionic component vanishes, $\xi \rightarrow 0$, the Hamiltonian in superspace \eqref{hamiltonian_in_superspace} reduces to the standard Hamiltonian of the KdV equation in space-time. A similar reduction can also be observed at the Lagrangian level, providing an additional consistency check.

\section{Summary and Discussion}
In this work, we constructed the constrained Hamiltonian formulation of the supersymmetric KdV system for the nontrivial case of $a=2$. Due to the degenerate nature of the Lagrangian for sKdV-2 \eqref{Lagrangian_skdv-2}, the Hamiltonian cannot be constructed using the usual Legendre transformation, instead, it must be constructed via the Dirac–Bergmann algorithm. The algorithm leads to a nontrivial secondary constraint, $\psi - \xi_x = 0$, which indicates a correlation between some degrees of freedom arising from the hidden structural restriction of the supersymmetric integrable system for fermionic fields. The constraint dynamics also result in nonlocality in the form of the inverse derivative operator in the constructed total Hamiltonian density \eqref{total_hamiltonian_density}. This nonlocality reveals that, in this formulation, $\xi_x$, rather than $\xi$, is the fundamental physical variable. The nonlocality does not appear in the superspace Hamiltonian, as it is removed through the relations between the fermionic component fields. The superspace Hamiltonian \eqref{hamiltonian_in_superspace} obtained in this study agrees with the previously constructed Hamiltonian expression of the sKdV-2 system~\cite{Mathieu1988}, up to the redefinition of the superfield associated with the introduction of the velocity potential. 

Our results show the role of constraints in the Hamiltonian construction of supersymmetric integrable systems and establish a bridge between component-level and superspace formulations. The approach presented here can be used for the investigation of the constrained Hamiltonian structures of supersymmetric formulations. One of the examples include nonlinear Schrödinger equations (NLSE). The supersymmetric extension of the NLSE, compatible with the Hamiltonian in superspace form, is given in~\cite{Ashok1995} as
\begin{align}\label{snlse 1}
    &i q_t +q_{xx} -2k(q^*q)q-2kq^*\phi_x\phi+2kq \phi \phi_x^*=0,
    \\ \label{snlse 2}
    &i\phi_t +\phi_{xx}-2kq^*q\phi=0.
\end{align}
Since the Hamiltonian structure of this extension shows notable similarities with the Hamiltonian structure of the supersymmetric extension of KdV, it is natural to consider this system in the future by using our approach.

In addition to the case of the supersymmetric NLSE, there exist nonlocal nonlinear integrable partial differential equations \cite{ablowitz2013integrable, ablowitz2017integrable,gurses2020,doikou2021}, often studied within the framework of PT symmetry, as well as nonlinear PDEs with delay \cite{polyanin2021construction,polyanin2024nonlinear,polyanin2025exact}, and integrable negative-order nonlinear wave equations \cite{verosky1991negative,qiao2012negative,ismailov2025inverse}, for which our construction can be naturally extended.

\section*{Acknowledgements}

\noindent We would like to thank the participants of the Symmetry and Geometry Workshop dedicated to the memory of Yavuz Nutku (Istanbul, 5-6 March 2026), where parts of this work were discussed. The work of Ilmar Gahramanov was partially supported by the Boğaziçi University Research Fund.

\appendix

\section{Dirac-Bergmann Algorithm}
Here, we briefly summarize the Dirac–Bergmann algorithm for constrained systems~\cite{Sundermeyer1982,Sardanashvily1995,Rothe2010,Deriglazov2010,salisbury2017,Lusanna2018,russkov2026}. This algorithm is a set of well-defined rules to construct the Hamiltonian from a degenerate Lagrangian. For a field $\phi(\textbf{r},t)$, a Lagrangian 
\begin{equation} \label{GenLag}
    L= \int d\textbf{r}\, \mathcal{L}\left[\phi_i, \nabla\phi_i, \dot\phi_i\right] 
\end{equation}
is called degenerate if the determinant of the Hessian matrix of the Lagrangian density $\mathcal{L}$
\begin{equation}
    \left\vert
    \frac{\delta^2 \mathcal{L}}{\delta {\dot{\phi}_i} \delta {\dot{\phi}_j}} 
    \right\vert=0\,,
\end{equation} 
vanishes. In such cases, the canonical momenta
\begin{equation}
\Pi_i = \frac{\partial \mathcal{L}}{\partial \dot{\phi}_i}
\end{equation}
do not uniquely determine the velocities.  
 
The Dirac-Bergmann algorithm introduces an initial set of primary constraints $c_i(\phi,\Pi)$ depending on the order of the degeneracy in the system. The total Hamiltonian density is constructed by adding the primary constraints with Lagrange multipliers to the canonical Hamiltonian density $\mathcal{H}_L$.
\begin{equation}
\mathcal{H}_T = \mathcal{H}_L + \lambda_i c_i.
\end{equation}
However, the dynamics of these constraints, $\{c_i,H\} = 0$, may also introduce a set of secondary constraints $\tilde{c}_i$. These constraints with their corresponding multipliers $\tilde{\lambda}_i$ also contribute to the total Hamiltonian. Thus, the new constraint's Hamiltonian density can be written as
\begin{equation}
    \mathcal{H}_c =\sum_i \left(\lambda_i c_i + \tilde{\lambda}_i \tilde{c}_i\right).
\end{equation}
This procedure is iterated until closure, meaning that all Lagrange multipliers are determined, to find the correct total Hamiltonian.

\bibliography{NLE}

@article{Nicolas2018,
  title = {The Korteweg-de Vries Equation},
  author = {Schalch, Nicolas},
  journal = {Proseminar: Algebra, Topology and Group Theory in Physics},
  year={2018}
}

@article{Nutku2002,
  title={Multi-Lagrangians for integrable systems},
  author={Nutku, Yavuz and Pavlov, MV},
  journal={Journal of Mathematical Physics},
  volume={43},
  number={3},
  pages={1441-1459},
  year={2002},
  publisher={American Institute of Physics}
}

@article{Nutku1984,
  title={Hamiltonian formulation of the KdV equation},
  author={Nutku, Yavuz},
  journal={Journal of Mathematical Physics},
  volume={25},
  number={6},
  pages={2007-2008},
  year={1984},
  doi={10.1063/1.526395}
}

@inproceedings{Nutku2001,
    author = "Nutku, Y.",
    title = "{Lagrangian approach to integrable systems yields new symplectic structures for KdV}",
    booktitle = "{NATO Advanced Research Workshop on Integrable Hierarchies and Modern Physical Theories (NATO ARW - UIC 2000)}",
    eprint = "hep-th/0011052",
    archivePrefix = "arXiv",
    month = "11",
    year = "2000"
}

@book{Deriglazov2010,
    author = {Deriglazov, A. A.},
    title = {Classical mechanics, Hamiltonian and Lagrangian formalism},
    year = {2010},
    publisher={Springer}
}

@article{Lusanna2018,
   title={Dirac–Bergmann constraints in physics: Singular Lagrangians, Hamiltonian constraints and the second Noether theorem},
   volume={15},
   number={10},
   journal={International Journal of Geometric Methods in Modern Physics},
   publisher={World Scientific Pub Co Pte Lt},
   author={Lusanna, Luca},
   year={2018},
   month={Oct},
   pages={1830004}
}

@book{Sardanashvily1995,
  title={Generalized Hamiltonian formalism for field theory: constraint systems},
  author={Sardanashvily, Gennadi A},
  year={1995},
  publisher={World Scientific}
}

@book{Sundermeyer1982,
  title={Constrained dynamics with applications to Yang-Mills theory, general relativity, classical spin, dual string model},
  author={Sundermeyer, Kurt},
  year={1982},
  publisher={Springer-Verlag}
}

@article{Mathieu1988,
    author = {Mathieu, Pierre},
    title = {Supersymmetric extension of the Korteweg–de Vries equation},
    journal = {Journal of Mathematical Physics},
    volume = {29},
    number = {11},
    pages = {2499-2506},
    year = {1988},
    month = {11},
    issn = {0022-2488},
    doi = {10.1063/1.528090},
    url = {https://doi.org/10.1063/1.528090},
}

@article{Ashok1995,
    author = {Brunelli, J. C. and Das, Ashok},
    title = {Tests of integrability of the supersymmetric nonlinear Schrödinger equation},
    journal = {Journal of Mathematical Physics},
    volume = {36},
    number = {1},
    pages = {268-280},
    year = {1995},
    month = {01},
    issn = {0022-2488},
    doi = {10.1063/1.531370},
    url = {https://doi.org/10.1063/1.531370}
}

@article{Pazarci2023,
title = {Hamiltonian formalism for nonlinear Schrödinger equations},
journal = {Communications in Nonlinear Science and Numerical Simulation},
volume = {121},
pages = {107191},
year = {2023},
issn = {1007-5704},
doi = {https://doi.org/10.1016/j.cnsns.2023.107191},
url = {https://www.sciencedirect.com/science/article/pii/S1007570423001090},
author = {Ali Pazarci and Umut Can Turhan and Nader Ghazanfari and Ilmar Gahramanov}
}

@article{Kupershmidt:1984pr,
    author = "Kupershmidt, B. A.",
    title = "{A SUPER KORTEWEG-DE VRIES EQUATION: AN INTEGRABLE SYSTEM}",
    doi = "10.1016/0375-9601(84)90693-5",
    journal = "Phys. Lett. A",
    volume = "102",
    pages = "213--215",
    year = "1984"
}

@article{Manin:1985hu,
    author = "Manin, Yu. I. and Radul, A. O.",
    title = "{A Supersymmetric extension of the Kadomtsev-Petviashvili hierarchy}",
    doi = "10.1007/BF01211044",
    journal = "Commun. Math. Phys.",
    volume = "98",
    pages = "65--77",
    year = "1985"
}

@article{fadeev1971a,
    author = "Zakharov, V. E. and Faddeev, L. D.",
    title =    "Korteweg-de Vries equation: A completely integrable Hamiltonian system",
    journal ="Functional Analysis and Its Applications" ,
    volume="5",
    pages="280-287",
    year = "1971",
    doi ="10.1007/BF01086739"
}

@Inbook{Flaschka1991,
author="Flaschka, H.
and Newell, A. C.
and Tabor, M.",
editor="Zakharov, Vladimir E.",
title="Integrability",
bookTitle="What Is Integrability?",
year="1991",
publisher="Springer Berlin Heidelberg",
address="Berlin, Heidelberg",
pages="73--114",
isbn="978-3-642-88703-1",
doi="10.1007/978-3-642-88703-1_3"
}

@article{Miki1976,
    author = {Kodama, Yuji and Wadati, Miki},
    title = {Theory of Canonical Transformations for Nonlinear Evolution Equations. I},
    journal = {Progress of Theoretical Physics},
    volume = {56},
    number = {6},
    pages = {1740-1755},
    year = {1976},
    month = {12},
    issn = {0033-068X},
    doi = {10.1143/PTP.56.1740}
}

@inbook{Newellch4,
author= {Alan C. Newell},
title = {The $\tau$-Function, the Hirota Method, the Painlevé Property and Bäcklund Transformations for the Korteweg—deVries Family of Soliton Equations},
booktitle = {Solitons in Mathematics and Physics},
chapter = {4},
publisher={Society for Industrial and Applied Mathematics},
pages = {113-144},
year={1985},
editor={},
doi = {10.1137/1.9781611970227.ch4}
}

@article{Lax1968,
author = {Lax, Peter D.},
title = {Integrals of nonlinear equations of evolution and solitary waves},
journal = {Communications on Pure and Applied Mathematics},
volume = {21},
number = {5},
pages = {467-490},
doi = {https://doi.org/10.1002/cpa.3160210503},
year = {1968}
}

@article{marchenko0a,
title = {The periodic Korteweg–de Vries problem},
journal = {Math. USSR-Sb.},
volume = {24},
number = {3},
pages = {319--344},
year = {1974},
doi = {https://doi.org/10.1070/SM1974v024n03ABEH002189},
author = {V. A. Marchenko}
}

@article{dubrovin19740a1,
title = {A periodicity problem for the Korteweg–de Vries and Sturm–Liouville equations. Their connection with algebraic geometry},
journal = {Dokl. Akad. Nauk SSSR},
volume = {219},
pages = {531-534},
year = {1974},
doi = {https://www.mathnet.ru/eng/dan38673},
author = { B. A. Dubrovin and S. P. Novikov}
}

@article{dubrovin19740a2,
title = {Periodic and conditionally periodic analogs of the many-soliton solutions of the Korteweg-de Vries equation},
journal = {Zh. Eksp. Teor. Fiz.},
volume = {67},
pages = {2131-2144},
year = {1974},
author = { B. A. Dubrovin and S. P. Novikov}
}

@book{Dickey2003,
author = {Dickey, L A},
title = {Soliton Equations and Hamiltonian Systems},
publisher = {World Scientific},
year = {2003},
doi = {10.1142/5108},
edition   = {2nd}
}

@article{Kruskal1965,
  title = {Interaction of "Solitons" in a Collisionless Plasma and the Recurrence of Initial States},
  author = {Zabusky, N. J. and Kruskal, M. D.},
  journal = {Phys. Rev. Lett.},
  volume = {15},
  issue = {6},
  pages = {240--243},
  numpages = {0},
  year = {1965},
  month = {Aug},
  publisher = {American Physical Society},
  doi = {10.1103/PhysRevLett.15.240}
}

@article{Zabusky_Galvin_1971, title={Shallow-water waves, the Korteweg-deVries equation and solitons}, volume={47}, DOI={10.1017/S0022112071001393}, number={4}, journal={Journal of Fluid Mechanics}, author={Zabusky, N. J. and Galvin, C. J.}, year={1971}, pages={811–824}}

@article{Morikawa1969,
    author = {Kever, H. and Morikawa, G. K.},
    title = {Korteweg‐de Vries Equation for Nonlinear Hydromagnetic Waves in a Warm Collision‐Free Plasma},
    journal = {The Physics of Fluids},
    volume = {12},
    number = {10},
    pages = {2090-2093},
    year = {1969},
    month = {10},
    issn = {0031-9171},
    doi = {10.1063/1.1692315}
}

@inbook{Whithamch16,
author={G. B. Whitham},
publisher = {John Wiley \& Sons, Ltd},
isbn = {9781118032954},
title = {Applications of the Nonlinear Theory},
booktitle = {Linear and Nonlinear Waves},
chapter = {16},
pages = {533-576},
doi = {https://doi.org/10.1002/9781118032954.ch16},
year = {1999}
}

@Inbook{Huang2005,
author="Huang, Guoxiang",
editor="Grimshaw, Roger",
title="Nonlinear Amplitude Equations and Soliton Excitations in Bose-Einstein Condensates",
bookTitle="Nonlinear Waves in Fluids: Recent Advances and Modern Applications",
year="2005",
publisher="Springer Vienna",
address="Vienna",
pages="169--196",
isbn="978-3-211-38025-3",
doi="10.1007/3-211-38025-6_6"
}

@article{ANTONOWICZ198795,
title = {A family of completely integrable multi-hamiltonian systems},
journal = {Physics Letters A},
volume = {122},
number = {2},
pages = {95-99},
year = {1987},
issn = {0375-9601},
doi = {https://doi.org/10.1016/0375-9601(87)90783-3},
url = {https://www.sciencedirect.com/science/article/pii/0375960187907833},
author = {Marek Antonowicz and Allan P. Fordy}
}

@article{Gardner-Kruskal-Miura-1967,
  title = {Method for Solving the Korteweg-deVries Equation},
  author = {Gardner, Clifford S. and Greene, John M. and Kruskal, Martin D. and Miura, Robert M.},
  journal = {Phys. Rev. Lett.},
  volume = {19},
  issue = {19},
  pages = {1095--1097},
  numpages = {0},
  year = {1967},
  month = {Nov},
  publisher = {American Physical Society},
  doi = {10.1103/PhysRevLett.19.1095}
}

@article{Zakharov-Shabat-1974,
    author = "Zakharov, V. E. and Shabat, A. B.",
    title = " A scheme for integrating the nonlinear equations of mathematical physics by the method of the inverse scattering problem. I",
    journal ="Functional Analysis and Its Applications" ,
    year = "1974",
    volume="8",
    pages="226-235",
    doi="10.1007/BF01075696"
}

@article{polyanin2024nonlinear,
  title={Nonlinear Schr{\"o}dinger equations with delay: Closed-form and generalized separable solutions},
  author={Polyanin, Andrei D and Kudryashov, Nikolay A},
  journal={Contemporary Mathematics},
  volume={5},
  number={4},
  pages={5783--5794},
  year={2024}
}

@article{ablowitz2017integrable,
  title={Integrable nonlocal nonlinear equations},
  author={Ablowitz, Mark J and Musslimani, Ziad H},
  journal={Studies in Applied Mathematics},
  volume={139},
  number={1},
  pages={7--59},
  year={2017},
  publisher={Wiley Online Library}
}

@article{ablowitz2013integrable,
  title={Integrable nonlocal nonlinear Schr{\"o}dinger equation},
  author={Ablowitz, Mark J and Musslimani, Ziad H},
  journal={Physical review letters},
  volume={110},
  number={6},
  pages={064105},
  year={2013},
  publisher={APS}
}

@article{polyanin2021construction,
  title={Construction of exact solutions to nonlinear PDEs with delay using solutions of simpler PDEs without delay},
  author={Polyanin, Andrei D and Sorokin, Vsevolod G},
  journal={Communications in Nonlinear Science and Numerical Simulation},
  volume={95},
  pages={105634},
  year={2021},
  publisher={Elsevier}
}

@article{polyanin2025exact,
  title={Exact solutions and reductions of nonlinear Schr{\"o}dinger equations with delay},
  author={Polyanin, Andrei D and Kudryashov, Nikolay A},
  journal={Journal of Computational and Applied Mathematics},
  volume={462},
  pages={116477},
  year={2025},
  publisher={Elsevier}
}

@inbook{Donagi_2003, place={Cambridge}, series={London Mathematical Society Lecture Note Series}, title={Geometry and integrability}, booktitle={Geometry and Integrability}, publisher={Cambridge University Press}, author={Donagi, Ron Y.}, editor={Mason, Lionel and Nutku, YavuzEditors}, year={2003}, pages={21–59}, collection={London Mathematical Society Lecture Note Series},doi={https://doi.org/10.1017/CBO9780511543135.004}}

@article{nutku1990,
    author = "Nutku, Y. and Oguz, O.",
    title =    " Bi-Hamiltonian structure of a pair of coupled KdV equations",
    journal =  "Il Nuovo Cimento B (1971-1996)",
    year = "1990",
    doi="10.1007/BF02742693"
}

@article{nutku_1985,
    author = {Nutku, Y.},
    title = {On a new class of completely integrable nonlinear wave equations. I. Infinitely many conservation laws},
    journal = {Journal of Mathematical Physics},
    volume = {26},
    number = {6},
    pages = {1237-1242},
    year = {1985},
    month = {06},
    issn = {0022-2488},
    doi = {10.1063/1.526530}
}

@article{nutku_1987,
    author = {Nutku, Y.},
    title = {On a new class of completely integrable nonlinear wave equations. II. Multi‐Hamiltonian structure},
    journal = {Journal of Mathematical Physics},
    volume = {28},
    number = {11},
    pages = {2579-2585},
    year = {1987},
    month = {11},
    issn = {0022-2488},
    doi = {10.1063/1.527749}
}

@article{nutku-gumral-1990,
    author = {Gümral, H. and Nutku, Y.},
    title = {Multi‐Hamiltonian structure of equations of hydrodynamic type},
    journal = {Journal of Mathematical Physics},
    volume = {31},
    number = {11},
    pages = {2606-2611},
    year = {1990},
    month = {11},
    issn = {0022-2488},
    doi = {10.1063/1.529012}
}

@article{YNutku_1983,
doi = {10.1088/0305-4470/16/18/020},
year = {1983},
month = {dec},
publisher = {},
volume = {16},
number = {18},
pages = {4195},
author = {Y Nutku},
title = {Canonical formulation of shallow water waves},
journal = {Journal of Physics A: Mathematical and General}
}

@article{delduc1996,
    author = {Delduc, F. and Ivanov, E. and Krivonos, S.},
    title = {N=4 super KdV hierarchy in N=4 and N=2 superspaces},
    journal = {Journal of Mathematical Physics},
    volume = {37},
    number = {3},
    pages = {1356-1381},
    year = {1996},
    month = {03},
    issn = {0022-2488},
    doi = {10.1063/1.531796}
}

@article{bonora1997,
author = {Bonora, L. and Krivonos, S.},
title = {Hamiltonian Structure and Coset Construction of the Supersymmetric Extensions of N=2 KdV Hierarchy},
journal = {Modern Physics Letters A},
volume = {12},
number = {39},
pages = {3037-3049},
year = {1997},
doi = {10.1142/S0217732397003162}}

@article{Nucci_2008,
doi = {10.1088/0031-8949/78/06/065011},
year = {2008},
month = {dec},
publisher = {},
volume = {78},
number = {6},
pages = {065011},
author = {Nucci, M C and Leach, P G L},
title = {The Jacobi Last Multiplier and its applications in mechanics},
journal = {Physica Scripta}
}

@article{prince2021,
  title = {The inverse problem in the calculus of variations: new developments},
  author = {Do, Thoan and Prince, Geoff},
  journal = {Communications in Mathematics},
  publisher = {University of Ostrava},
  volume = {29},
  number = {1},
  pages = {131 - 149},
  year = {2021},
  doi = {10.2478/cm-2021-0008},
}

@article{montesinos2026,
author = {Montesinos, Merced and Gonzalez, Diego and Meza, Jorge},
title = {Combining Symmetries and Helmholtz’s Conditions to Construct Lagrangians},
journal = {Advances in Mathematical Physics},
volume = {2026},
number = {1},
pages = {9534805},
doi = {https://doi.org/10.1155/admp/9534805},
year = {2026}
}

@article{nutku-gurses-1981,
    author = {Gürses, Metin and Nutku, Yavuz},
    title = {New nonlinear evolution equations from surface theory},
    journal = {Journal of Mathematical Physics},
    volume = {22},
    number = {7},
    pages = {1393-1398},
    year = {1981},
    month = {07},
    issn = {0022-2488},
    doi = {10.1063/1.525079}
}

@article{popowicz2019,
    author = "Popowicz, Ziemowit",
    title = "N = 2 Supercomplexification of the Korteweg-de Vries, Sawada-Kotera and Kaup-Kupershmidt Equations",
    journal = "Journal of Nonlinear Mathematical Physics",
    volume="26",
    year = "2019",
    doi="10.1080/14029251.2019.1591732"
}

@article{das-popowicz-2000,
title = {New nonlocal charges in SUSY-B integrable models},
journal = {Physics Letters A},
volume = {274},
number = {1},
pages = {30-36},
year = {2000},
issn = {0375-9601},
doi = {https://doi.org/10.1016/S0375-9601(00)00523-5},
author = {Ashok Das and Ziemowit Popowicz}
}

@article{doikou2021,
title = {Grassmannian flows and applications to non-commutative non-local and local integrable systems},
journal = {Physica D: Nonlinear Phenomena},
volume = {415},
pages = {132744},
year = {2021},
issn = {0167-2789},
doi = {https://doi.org/10.1016/j.physd.2020.132744},
author = {Anastasia Doikou and Simon J.A. Malham and Ioannis Stylianidis},
}

@misc{rybkin2026,
      title={On the inverse scattering transform for the KdV equation with summable initial data}, 
      author={Alexei Rybkin},
      year={2026},
      eprint={2604.14412},
      archivePrefix={arXiv},
      primaryClass={math-ph},
      url={https://arxiv.org/abs/2604.14412}, 
}

@book{lannes2013,
    author = "David Lannes",
    title = "The Water Waves Problem: Mathematical Analysis and Asymptotics",
    publisher="Mathematical Surveys and Monographs, American Mathematical Society",
    volume="188", doi="https://doi.org/10.1090/surv/188",
    year = "2013"
}

@article{grudsky2023,
author = {Grudsky, Sergei M. and Kravchenko, Vladislav V. and Torba, Sergii M.},
title = {Realization of the inverse scattering transform method for the Korteweg–de Vries equation},
journal = {Mathematical Methods in the Applied Sciences},
volume = {46},
number = {8},
pages = {9217-9251},
doi = {https://doi.org/10.1002/mma.9049},
year = {2023}
}

@article{Ma2022,
title = {N-soliton solutions and the Hirota conditions in (1 + 1)-dimensions},
author = {Wen Xiu Ma},
pages = {123--133},
volume = {23},
number = {1},
journal = {International Journal of Nonlinear Sciences and Numerical Simulation},
doi = {doi:10.1515/ijnsns-2020-0214},
year = {2022},
}

@article{gurses2020,
title = {Nonlocal KdV equations},
journal = {Physics Letters A},
volume = {384},
number = {35},
pages = {126894},
year = {2020},
issn = {0375-9601},
doi = {https://doi.org/10.1016/j.physleta.2020.126894},
author = {Metin Gürses and Aslı Pekcan},
}

@book{Rothe2010,
    author = "Rothe, Heinz J. and Rothe, Klaus D.",
    title = "{Classical and quantum dynamics of constrained Hamiltonian systems}",
    publisher = "World Scientific",
    year = "2010"
}

@article{salisbury2017,
    author = "Salisbury, Donald and Sundermeyer, Kurt",
    title =    "Léon Rosenfeld’s general theory of constrained Hamiltonian dynamics" ,
    journal = "The European Physical Journal H",
    year = "2017",
    doi="10.1140/epjh/e2016-70042-7",
    volume="42"
}

@misc{russkov2026,
      title={Remarks on Dirac-Bergmann algorithm, Dirac's conjecture and the extended Hamiltonian}, 
      author={Kirill Russkov},
      year={2026},
      eprint={2602.00284},
      archivePrefix={arXiv},
      primaryClass={hep-th},
      url={https://arxiv.org/abs/2602.00284}, 
}

@article{filiz2018,
    author = {Çağatay Uçgun, Filiz and Esen, Oğul and Gümral, Hasan},
    title = {Reductions of topologically massive gravity I: Hamiltonian analysis of second order degenerate Lagrangians},
    journal = {Journal of Mathematical Physics},
    volume = {59},
    number = {1},
    pages = {013510},
    year = {2018},
    month = {01},
    issn = {0022-2488},
    doi = {10.1063/1.5021948},
}

@Article{ma2024,
AUTHOR = {Ma, Hongcai and Qi, Xinru and Deng, Aiping},
TITLE = {Exact Soliton Solutions to the Variable-Coefficient Korteweg–de Vries System with Cubic–Quintic Nonlinearity},
JOURNAL = {Mathematics},
VOLUME = {12},
YEAR = {2024},
NUMBER = {22},
ARTICLE-NUMBER = {3628},
ISSN = {2227-7390},
DOI = {10.3390/math12223628}
}

@article{Verheest_Hereman_2025, title={The Gardner equation and acoustic solitary waves in plasmas}, volume={91}, DOI={10.1017/S0022377825100615}, number={4}, journal={Journal of Plasma Physics}, author={Verheest, Frank and Hereman, Willy A.}, year={2025}, pages={E117}}

@misc{sonja2024,
      title={Selected aspects of the Korteweg-de Vries equation}, 
      author={Sonja Hohloch and Federico Zadra},
      year={2024},
      eprint={2411.18504},
      archivePrefix={arXiv},
      primaryClass={nlin.SI},
      url={https://arxiv.org/abs/2411.18504}, 
}

@article{ismailov2025inverse,
  title={Inverse scattering method via the Gel’fand--Levitan--Marchenko equation for some negative-order nonlinear wave equations},
  author={Ismailov, Mansur I and Sabaz, Cihan},
  journal={Theoretical and Mathematical Physics},
  volume={222},
  number={1},
  pages={20--33},
  year={2025},
  publisher={Springer}
}

@article{qiao2012negative,
  title={Negative-order Korteweg--de Vries equations},
  author={Qiao, Zhijun and Fan, Engui},
  journal={Physical Review E—Statistical, Nonlinear, and Soft Matter Physics},
  volume={86},
  number={1},
  pages={016601},
  year={2012},
  publisher={APS}
}

@article{verosky1991negative,
  title={Negative powers of Olver recursion operators},
  author={Verosky, John M},
  journal={Journal of mathematical physics},
  volume={32},
  number={7},
  pages={1733--1736},
  year={1991},
  publisher={American Institute of Physics}
}

@article{gumral2022,
author = {G\"{u}mral, Hasan},
title = {Dirac’s analysis and Ostrogradskii’s theorem for a class of second-order degenerate Lagrangians},
journal = {International Journal of Geometric Methods in Modern Physics},
volume = {19},
number = {01},
pages = {2250008},
year = {2022},
doi = {10.1142/S0219887822500086}}

\end{document}